\pdfoutput=1
\documentclass[11pt]{article}

\usepackage[]{emnlp2022}

\usepackage{times}
\usepackage{latexsym}

\usepackage[T1]{fontenc}
\usepackage[utf8]{inputenc}

\usepackage{microtype}

\usepackage{inconsolata}
\usepackage{hyperref}       % hyperlinks
\usepackage{url}            % simple URL typesetting
\usepackage{booktabs}       % professional-quality tables
\usepackage{amsfonts}       % blackboard math symbols
\usepackage{nicefrac}       % compact symbols for 1/2, etc.
\usepackage{microtype}      % microtypography
\usepackage{xcolor}    
\usepackage{amsmath}
\usepackage{subfig}
\usepackage{enumitem}
\usepackage{graphicx}
 \usepackage{multirow}
 \usepackage{arydshln}

\title{Detect-Localize-Repair: A Unified Framework for\\Learning to Debug with CodeT5}

\author{Nghi D. Q. Bui, Yue Wang, Steven C.H. Hoi\\
	 Salesforce Research Asia \\
	\texttt{\{nghi.bui,wang.y,shoi\}@salesforce.com}\\}

\begin{document}
\maketitle
%!TEX root=emnlp2022.tex
% Automated software debugging is an important task for increasing the productivity of software developers. The use of neural-based techniques for debugging-related tasks such as bug localization and program repair is a recent trend (or code refinement). However, recent techniques are aimed at either bug localization or program repair. In this paper, we propose a new unified framework based on a pretrained language model that can perform both tasks. We propose a multi-modal neural architecture with three objectives: a bug detection objective to determine whether a given code snippet is buggy or not, a line ranking objective to rank the buggy lines according to some scores, and a repair goal to translate the buggy code to correct code.

\begin{abstract}
Automated software debugging is a crucial task for improving the productivity of software developers. Many neural-based techniques have been proven effective for debugging-related tasks such as bug localization and program repair (or bug fixing). However, these techniques often focus only on either one of them or approach them in a stage-wise manner, ignoring the mutual benefits between them. In this work, we propose a novel unified \emph{Detect-Localize-Repair} framework  based on a pretrained programming language model CodeT5 to seamlessly address these tasks, named CodeT5-DLR. Specifically, we propose three objectives to adapt the generic CodeT5 for debugging: a bug detection objective to determine whether a given code snippet is buggy or not, a bug localization objective to identify the buggy lines, and a program repair objective to translate the buggy code to its fixed version. 
We evaluate it on each of these tasks and their combined setting on two newly collected line-level debugging datasets in Java and Python. Extensive results show that our model significantly outperforms  existing baselines from both NLP and software engineering domains.
% Further analysis reveals that these  objectives  can complement to each other and benefit each task (bug detection, bug localization, and program repair) compared to models trained only on a single task. 
\end{abstract}

\section{Introduction}
\label{sec:intro}
Program debugging is crucial, yet most cost-dominating in software development. The goal of program debugging is to localize erroneous lines of a program (\emph{bug localization}) and fixes this buggy patch (\emph{program repair}).
The majority of debugging tools falls into two categories: program analysis-based and neural-based. To debug a program, program analysis-based techniques employ compiler-based and software engineering theory to build code analysis tools. These methods have a significant disadvantage in that they are not scalable to large and complicated programs.
On the other hand, a recent trend is to use neural-based techniques~\cite{lutellier2020coconut, jiang2021cure,zhu2021syntax,mashhadi2021applying,ding2020patching, wang2021codet5}  based on the \textit{naturalness hypothesis} of software code~\cite{hindle2016naturalness}. 
They adopt a generic data-driven approach to train neural networks to automatically acquire bug-fix patterns through learning from a massive corpora of previous bug-fixes.

% These techniques rely on large code corpora that can be used as training and test sets, allowing machine learning methods to learn and probabilistically reason about coding practice on a large scale. The goal is to develop useful tools on top of these models  and program debugging is one of the topics that has received a lot of attention in order to employ such neural-based model to build such tools. 

However, these techniques suffer from a few major drawbacks.
First, they often utilize code-specific or language-specific features such as control flow, data flow,  and abstract syntax trees (ASTs), which requires a significant amount of engineering effort for a careful design of code representations and thus hinders their applicability to more diverse domains or programming languages.
Second, recent studies have focused on detecting bugs at coarse-grained code granularity such as function level or file level, which has been shown to be impractical in real-world use~\cite{zou2019empirical}. It is also not ideal to localize bugs at too fine-grained level like the token level, which might lead to a large number of false positives~\cite{allamanis2021self}. Line-level or statement-level bug localization, on the other hand, has been extensively studied in the domain of program analysis, such as spectrum-based bug localization~\cite{abreu2009spectrum,le2013theory, xie2013theoretical,abreu2007accuracy}, in which the buggy statements are localized based on the signals of failed test cases, or mutation-based bug localization~\cite{jia2010analysis,papadakis2015metallaxis,zhang2018predictive}, in which the buggy statements are localized by randomly mutating the statements and measuring against a test suite. However, these traditional techniques necessitate the execution of test cases in order to complete the bug localization process, which has scalability issues.
As a result, we propose that line-level bug detection is more reasonable if we can use large-scale bug fixes datasets; and it corresponds to how human developers read and debug programs. Finally, these techniques are only intended for either bug localization or program repair, or treat them separately as two stages, which fails to exploit their potential mutual benefits. Intuitively, a preciser bug detector is able to inform the repairer with more accurate buggy information to aid the bug fixing, while a good repairer usually has strong code understanding that is also required in bug detection.

% To begin with, the majority of the techniques work by utilising unique source code features such as control flow, data flow, abstract syntax trees, and so on. However, to extract source code features, these techniques necessitate careful design of code representations and a significant amount of engineering effort, limiting their applicability to specific domains or programming languages. Second, recent studies have focused on detecting bugs at coarse-grained code granularity (function level or file level).
% According to recent surveys on the adoption of such tools in real-world use~\cite{zou2019empirical}, this is not a very useful signal for developers. Buggy signals should be provided by the tools at finer-grained code levels, such as the statement, line, or token level. The neural-based bug detection tool at the token level, on the other hand, is said to be too fine-grained and generate a large number of false positives, making the tool become impractical~\cite{allamanis2021self}. The line level is the most reasonable level of bug detection because it corresponds to how the developer reads and debugs programs. Finally, these techniques are only intended for bug localization or program repair, whereas a good program debugging tool should be capable of both.

To address these issues, we propose a unified framework for adapting a general pretrained programming language model for line-level debugging and repair. Our framework is based on a key observation about how programmers debug their code. First, he or she must determine whether or not a function is buggy. If it is buggy, the developer must localize the problematic line and provide a patch (repair). Inspired from this procedure, we propose
three fine-tuning objectives on top of a pretrained language model for debugging. 
Using  pretrained language models for code has two advantages. First, by treating code as natural language, it reduces the effort of the code representation engineering process. Second, it can leverage the pretrained knowledge gained from large number of source code. 
We employ CodeT5~\cite{wang2021codet5} as the foundation model which has achieved  state-of-the-art results on a wide range of code intelligence tasks. CodeT5 is pretrained on large-scale  code corpus collected from Github using code-aware objectives, which endows the model with strong code understanding capability. 

%The first objective is the function-level bug detection task, which is to detect if a given code contains any bug or not (D). Second, to provide more useful information for developer at finer-grained code granularity, we introduce a bug localization objective to identify the exact buggy lines in the code snippet (L). The third one is a program repair objective to translate the buggy code to its corresponding correct code (if any) (R). We expect these tasks can complement to each other and result in a strong and comprehensive program debugging tool that is capable of different debugging-related tasks. Using all of the objectives for fine-tuning result into a model called CodeT5-DLR.

The first objective is the function-level bug detection task, which entails determining whether or not a particular piece of code includes a bug (D). Second, in order to give developers with more valuable information at a finer-grained code granularity, we propose a bug localization aim to identify the exact lines of code that include bugs (L). The third ojective is the program repair, which is used to convert the buggy code to the correct code (if applicable) (R). We expect that these tasks will complement one another and culminate in a robust, all-encompassing software debugging tool capable of doing many debugging-related tasks. CodeT5-DLR is the model created by applying all of the fine-tuning objectives.

%Specifically, we propose three objectives to adapt the generic CodeT5 for debugging (CodeT5-DLR). The first objective is the function-level bug detection task, which is to detect if a given code contains any bug or not (D). Second, to provide more useful information for developer at finer-grained code granularity, we introduce a bug localization objective to identify the exact buggy lines in the code snippet (L). The third one is a program repair objective to translate the buggy code to its corresponding correct code (if any) (R). We expect these tasks can complement to each other and result in a strong and comprehensive program debugging tool that is capable of different debugging-related tasks.

% The main idea is to fine-tune CodeT5 on three objectives that are specifically designed for program debugging. The first objective is the function-level bug detection task, which is to detect if a given code contains any bug or not. Second, to provide more useful information to developer at finer-grained code granularity, we introduce a bug localization objective to identify the exact buggy lines in the code snippet. The third one is a program repair objective to translate the buggy code to its corresponding correct code (if any).
% ~\footnote{Patch generation refers to the process of creating the correct fix for a buggy code snippet.}. These 3 objectives complement on each other and each will assist the others to make the overall performance better.

To evaluate on the whole debugging procedure (D-L-R), we newly collected two large-scale bug-fix datasets in Java and Python programming languages from Github commits, which is released to facilitate future research as part of our contribution. We consider two types of bugs: single-line and multi-line.
We evaluate our CodeT5-DLR on three separate debugging-related tasks: function-level bug detection, line-level bug localization, and program repair.  Our evaluation results show that our model significantly  outperforms existing baselines on all of the tasks. We further conduct  ablation studies to demonstrate that jointly training with the three objectives  yields better performance than separately training on each single task. Finally, we design a unified evaluation task to combine all of these tasks that is to mimic how developers localize and fix bugs in real-world scenario, where our model is able to correctly localize 33.93\% buggy lines and repair 46.93\%  for a single line  bug fix Java dataset.

% We trained our model based on the datasets of bug-fixing code changes collected on Github. Once trained,
% We evaluate our models for 4 separate tasks: function-level bug detection, line-level bug detection, program repair and a unify program-debugging task.  Our evaluation shows that our model perform the best among the baselines for all of the tasks. We also conduct additional ablation studies to demonstrate that training the three objectives concurrently improves performance significantly more than training each task separately. Finally, we perform a unify evaluation task that mimic how developers localize and fix bugs. This is to evaluate how well our model performs in real-world scenario when a developer needs to debug a program.

Our major contributions are three-fold:
\begin{itemize}[leftmargin=*]
	\item We propose a unified Detect-Localize-Repair framework (CodeT5-DLR) based on CodeT5 to seamlessly solve three program debugging tasks: function-level bug detection, line-level bug localization and program repair.
	\item We introduce two newly collected large-scale line-level debugging datasets in Java and Python programming languages with useful information for future research, including the buggy line indicator; and the before-fixed version and the after-fixed version of code snippets.
	\item We conduct extensive evaluations on each of debugging tasks and their combined task, where our model outperforms  existing baselines with a significant margin.
\end{itemize}

\section{Method}
\label{sec:model}
\begin{figure*}[t]
	\centering
	\includegraphics[width=0.99\textwidth]{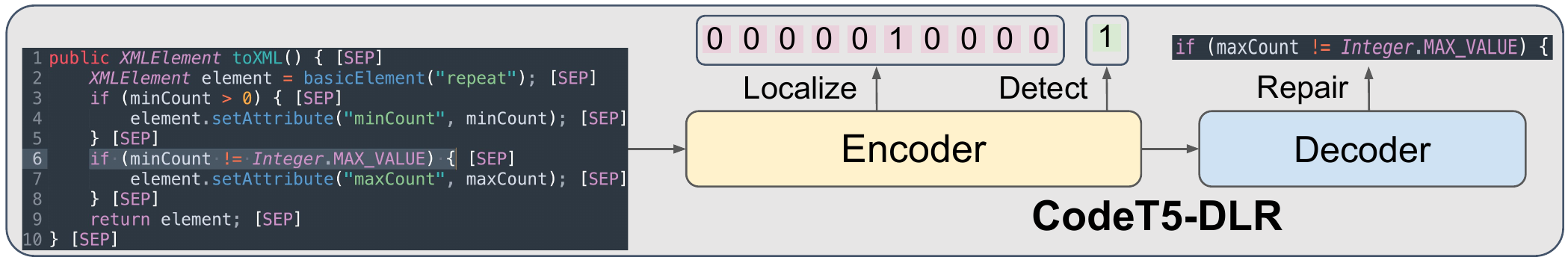}
	\caption{An overview of our CodeT5-DLR framework to jointly detect, localize, and repair bugs.}
	\label{fig:overview}
\end{figure*}

As shown in Figure~\ref{fig:overview}, we present a unified framework to jointly address three crucial tasks in program debugging: \emph{bug detection} (whether a given code snippet contains bugs), \emph{bug localization} (which lines are buggy), and \emph{program repair} (how to repair bugs).
We first define the input/output formulation of these tasks in \S~\ref{ssec:formulation} and then revisit the foundation model CodeT5 in \S~\ref{ssec:codet5}, followed by introducing each of tasks in \S~\ref{ssec:localize_repair} in detail.

% https://docs.google.com/presentation/d/1pkVoHaZzpjReKFpRIirc1jx50YrlKOAKxlallWOdYbw/edit?usp=sharing

%\subsection{Input and Output}
\subsection{Problem Definition}
\label{ssec:formulation}
Let $\mathcal{D}$ be a program debugging dataset consisting of $|\mathcal{D}|$  triplets of $(X,Y,F)$. $X$ is the source program patch at function level and $Y=\{y_1, ..., y_L\}$ is its buggy labels for each line, where $y_i \in [0,1]$ represents whether the $i$-th line is buggy or not and $L$ denotes the number of lines in $X$. $F$ is the target fixed program if source patch $X$ contains any buggy lines, otherwise it is an empty string. Let $y$ denote such function-level binary label and $y=1$ if there exists $y_i=1$ and else $y=0$.
For the input format of $X$, we insert a special token \texttt{[SEP]} for each line to inform the end of line information. 

\subsection{Revisiting CodeT5} \label{ssec:codet5}
CodeT5~\cite{wang2021codet5} is a unified pretrained encoder-decoder language models for code. It was pretrained on a large-scale source code corpus collected from Github which consists of 8 different programming languages (including Java and Python). Moreover, CodeT5 proposed an identifier-aware pretraining objective to endow the model with code-specific knowledge. Besides that, it employs a code-specific Byte-Pair Encoding (BPE)~\cite{DBLP:conf/acl/SennrichHB16a} tokenizer that is able to avoid Out-of-Vocabulary (OoV) problems. CodeT5 has achieved state-of-the-art performance on a wide range of code intelligence tasks in CodeXGLUE~\cite{lu2021codexglue} such as defect detection and code refinement. In this work, we adapt CodeT5 as our foundation model and propose a new unified framework to jointly solve bug localization and repair.

\subsection{Detect-Localize-Repair Framework} \label{ssec:localize_repair}
In this subsection, we introduce how to adapt CodeT5 for bug detection, localization, and repair. 

\paragraph{Function-Level Bug Detection} \label{ssec:fbd}
The goal of this task is to detect whether a function contains any bugs. Given an input code patch $X$, we aim to learn the binary probability of $P_{\theta}(y|X)$ with CodeT5 parameterized by $\theta$. Specifically, we pass the source patch $X$ to the encoder of CodeT5 and adopt the last encoder state as the sequence representation of $X$, followed by a linear layer on top of it for a binary classification. This task is optimized with a standard cross entropy loss (denoted as $\mathcal{L}_{detect}$) in training and a patch is considered as buggy if the predicted probability is higher than a threshold of $0.5$ in inference.

\paragraph{Line-Level Bug Localization} \label{ssec:lbd}
A further step of bug detection is to localize which exact lines are buggy. This is an important intermediate task for the final successful repair. This task is formulated to compute $P_{\phi}(Y|X)$, which $\phi$ denotes the parameter of the encoder of CodeT5. Specifically, we gather the last layer states of all \texttt{[SEP]} tokens from the encoder and map them to a vector of probabilities $\hat{Y}=\{\hat{y}_1, ..., \hat{y}_L\}$. We approach the bug localization as a sequence labeling task and apply the binary cross entropy loss (denoted as $\mathcal{L}_{localize}$) between $\hat{Y}$ and $Y$ during training. For inference, we obtain the top-$k$ predictions and measure how they match the ground truth with retrieval metrics. Notably, we consider two settings of bug localization where the source patch can  have only one single buggy line or multiple buggy lines.

\paragraph{Program Repair} \label{ssec:apr}
This task aims to translate a buggy source patch $X$ into its fixed version $F$. Formally, we aim to learn the probability
$P_{\theta}(F|X)=\prod_{j=1}^{n}P_{\theta}(Y_j| X,F_1:F_j-1)$, where $F_1:F_j-1$ is the previous sequence before the $j$-th token and $n$ denotes the number of tokens in the target sequence $F$. We approach this task a sequence-to-sequence problem and train with a standard sequence generation loss (denoted as $\mathcal{L}_{repair}$). During inference, we adopt beam search to generate a ranked list of fixed candidate patches.

\paragraph{Joint Training} During training, we adopt multi-task learning to simultaneously optimize these three tasks by combining their losses in an end-to-end manner: 
\begin{equation}
    \mathcal{L}_{all} = \mathcal{L}_{detect} + \mathcal{L}_{localize} + \mathcal{L}_{repair}
\end{equation}
The intuition behind this design is that these tasks are highly related and can  complement to each other. For instance, a preciser bug locator can better inform the repairer with bug location to aid the bug fixing. Therefore, we expect these tasks can benefit from such a joint training paradigm.

\section{Datasets}
\label{sec:dataset}

We collect two new datasets, one is the single line bug-fixes pair and the other is the multi-line bug-fixes pair. The single line bug-fixes~\cite{karampatsis2020often} have been considered recently as one of the major issues that affects the quality of source code. These bugs can be fixed easily with simple code changes such as changing operators, renaming identifiers, swaping variables and so on. However, these bugs occurred frequently, and current static-based techniques are incapable of detecting them accurately (less than $10\%$ in accuracy).
%fixes~\cite{karampatsis2020often, richter2022tssb}

Existing datasets~\cite{karampatsis2020often, richter2022tssb}, however, are not suitable for our purpose due to three reasons: \textbf{(1)} they contain only the code changes at the file level while our goal is to detect buggy lines at both function level and line level; \textbf{(2)} they does not contain the before and after function-level information of the code changes but  only the patches at line-level; and \textbf{(3)} these datasets are mostly for single-line bug fixes, while our goal is to extend for a more realistic setting of multi-line bugs. With such reasons, we decide to collect datasets by ourselves for a comprehensive evaluation.  

We follow similar steps from ~\cite{karampatsis2020often} to collect two datasets in Java and Python.
Concretely, we extract bug-fixes code changes from Github commits, we use Pydriller~\cite{Spadini2018}, a tool that mines software repositories from Git. To decide if a commit fixes a bug, we follow~\citet{karampatsis2020often}  to check if its commit message contains at least one of the
keywords: \textit{error}, \textit{bug}, \textit{fix}, \textit{issue}, \textit{mistake}, \textit{incorrect}, \textit{fault},
\textit{defect}, \textit{flaw}, and \textit{type}. This heuristic has been  shown to achieve $96\%$ accuracy on a set of 300 manually verified commits~\cite{ray2016naturalness} and 97.6\% on a set of 384 manually verified commits~\cite{tufano2018empirical}. 

The code changes are made up of three parts of a source file: before changes, after changes, and the difference between the two (patch). However, because we want to localise bugs at the \textit{function and line level} rather than the file level, we need to perform additional preprocessing to extract the code changes at the \textit{function level}. We use Lizard~\footnote{\url{https://github.com/terryyin/lizard}, a code analysis tool for extracting functions from source code files that supports multiple programming languages.} to extract the functions and compare the different between the functions from the before and after version of a source file (obtained from Pydriller). 
%Figure~\ref{fig:sample} depicts a sample from our data collection process that was successfully extracted.

We end up with two datasets of different types and languages, one is for single-line bug-fixes in Java (\textbf{SL-Java}) and the other is the multi-lines bug-fixes in Python (\textbf{ML-Python}).
For \textbf{SL-Java}, beside the code changes for bug fixes, we also follow~\citet{karampatsis2020often} to use tree-sitter~\footnote{\url{https://github.com/tree-sitter/tree-sitter}} to identify $13$ bug patterns for the single buggy lines. We do not aim to detect the exact patterns because they can be easily detected using matching rules on ASTs if we can localize if a line is buggy. However, the patterns is useful in analysing how well our techniques work on each pattern, allowing us to gain a deeper understanding of the debugging process.

Table~\ref{tab:bug_patterns_detail} shows the details of 13 bug patterns in our SL-Java dataset. We also provide a bug-fix sample for each of the pattern. Overall, the \texttt{CHANGE\_IDENTIFIER} appears the most and \texttt{SWAP\_BOOLEAN\_LITERAL} appears the least.
\begin{table*}[t]
	\centering
	\caption{Detail of 13 Bug Patterns in SL-Java dataset.}
	\label{tab:bug_patterns_detail}
	\resizebox{1.0\textwidth}{!} {
		\begin{tabular}{c|c|c|cc}
			\hline
			\multirow{2}{*}{Index} & \multirow{2}{*}{Bug Pattern}    & \multirow{2}{*}{Num Instances} & \multicolumn{2}{c}{Example}                                                                                                      \\ \cline{4-5} 
			&                                 &                                & \multicolumn{1}{c|}{Before}                               & Affter                                                               \\ \hline
			P0                     & CHANGE OPERATOR                 & 3784                           & \multicolumn{1}{c|}{logLevel\textgreater{}=Log.ASSERT}    & logLevel\textless{}=Log.ASSERT                                       \\ \hline
			P1                     & CHANGE\_OPERAND                 & 1089                           & \multicolumn{1}{c|}{x\textless{}=1}                       & z\textless{}=1                                                       \\ \hline
			P2                     & CHANGE\_IDENTIFIER              & 15006                          & \multicolumn{1}{c|}{this.userDn}                          & this.userName                                                        \\ \hline
			P3                     & CHANGE\_NUMERAL                 & 8969                           & \multicolumn{1}{c|}{player.stepHeight=0.5F}               & player.stepHeight=0.6F                                               \\ \hline
			P4                     & CHANGE\_CALLER\_IN\_FUNCTION    & 3180                           & \multicolumn{1}{c|}{mBlockStream.remaining()}             & inStream.remaining()                                                 \\ \hline
			P5                     & CHANGE\_UNARY\_OPERATOR         & 2212                           & \multicolumn{1}{c|}{!segment.isOk()}                      & segment.isOk()                                                       \\ \hline
			P6                     & OVERLOAD\_METHOD\_MORE\_ARGS    & 9732                           & \multicolumn{1}{c|}{Messaging.sendTr(sender,key)}         & Messaging.sendTr(sender,key,npc.getName())                           \\ \hline
			P7                     & OVERLOAD\_METHOD\_DELETED\_ARGS & 2577                           & \multicolumn{1}{c|}{registerCommandsNow(commands)}        & registerCommandsNow()                                                \\ \hline
			P8                     & DIFFERENT\_METHOD\_SAME\_ARGS   & 18914                          & \multicolumn{1}{c|}{server.getStartedLabel()}             & server.getStartedName()                                              \\ \hline
			P9                     & MORE\_SPECIFIC\_IF              & 3981                           & \multicolumn{1}{c|}{getIndex()\textgreater{}=arrayLength} & arrayLength\textgreater{}0 \&\& getIndex()\textgreater{}=arrayLength \\ \hline
			P10                    & LESS\_SPECIFIC\_IF              & 3919                           & \multicolumn{1}{c|}{pluginId==null}                       & pluginId==null || pluginID.length()==0                               \\ \hline
			P11                    & SWAP\_ARGUMENTS                 & 1244                           & \multicolumn{1}{c|}{new Duration(DateTime.now(),time)}    & new Duration(time, DateTime.now()                                    \\ \hline
			P12                    & SWAP\_BOOLEAN\_LITERAL          & 897                            & \multicolumn{1}{c|}{doTest(false)}                        & doTest(true)                                                         \\ \hline
		\end{tabular}
	}
\end{table*}

Table~\ref{tab:dataset} shows the statistics of our datasets, which have been splited into training,  validation, testing sets. For function-level bug detection task, we use the whole code snippet. The buggy or non-buggy label is decided by simply treating the before version as the buggy (label 0) and the after version as non-buggy (label 1). 
For line-level bug localization task, we use the buggy line's number information to train the model to localize which line is buggy. For program repair, the before version is used as the source input and the after version is used as the target sequence.

\begin{figure*}[t]
	\centering
	\includegraphics[width=1.0\textwidth]{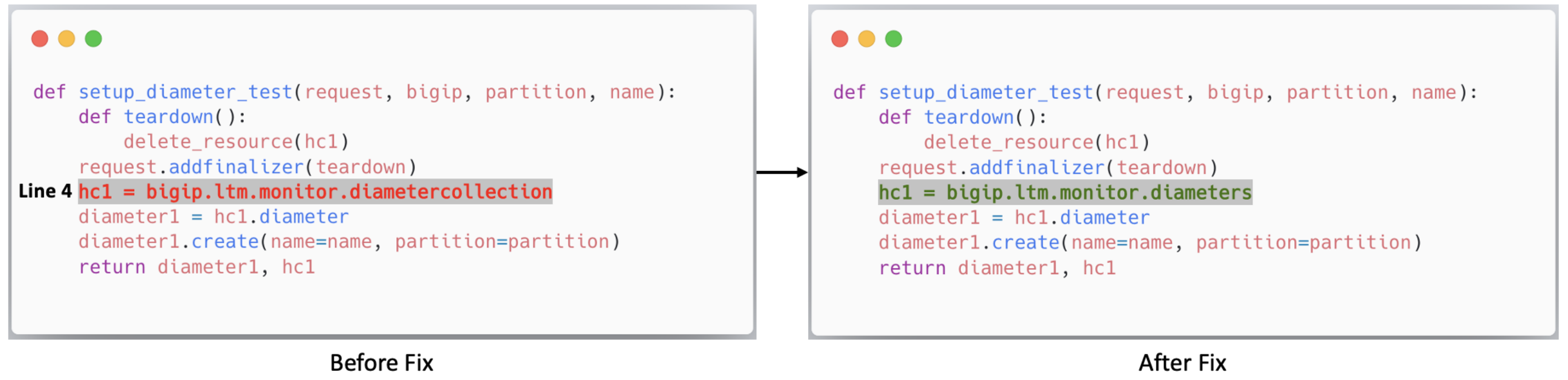}
	\caption{A sample of an instance in our SL-Java dataset}
	\label{fig:data_samples}
\end{figure*}

Figure~\ref{fig:data_samples} shows a sample in our SL-Java dataset. A sample comprises primarily of the buggy code snippet (Before) and the fixed code snippet (After) (After). It also includes the line number of the defective line. Multiple buggy lines are present in the ML-Python samples as opposed to a single buggy line. Each instance also include other meta data, such as commit message, commit id (SHA1) and project name so that it is easy to trace back the original bug information.

\begin{table}[t]
	\centering
	\caption{Statistics of our datasets}
	\label{tab:dataset}
	\resizebox{0.49\textwidth}{!} {
	\begin{tabular}{l|cc|cc}
		\hline
		\multirow{2}{*}{Split} & \multicolumn{2}{c|}{SL-Java}                      & \multicolumn{2}{c}{ML-Python}                     \\ \cline{2-5} 
		& \#Projects &  \multicolumn{1}{|c|}{\#Instances} & \#Projects &  \multicolumn{1}{|c}{\#Instances} \\ \hline
		Train             & \multicolumn{1}{c|}{2,348}         & 52,789         & \multicolumn{1}{c|}{4,428}         & 132,243        \\ \hline
			
		Val               & \multicolumn{1}{c|}{335}          & 7,465          & \multicolumn{1}{c|}{480}          & 22,395         \\ \hline
		Test              & \multicolumn{1}{c|}{700}          & 15,250         & \multicolumn{1}{c|}{867}         & 35,457         \\ \hline
	\end{tabular}
	}
\end{table}

\section{Experiments} \label{sec:exp}

We employ the CodeT5-base (220M) \footnote{\url{https://github.com/salesforce/CodeT5}} as the foundation model for our unified CodeT5-DLR framework. For the purpose of ablation study, we also consider three variants of our model: CodeT5-D trained with $\mathcal{L}_{detect}$, CodeT5-L trained with $\mathcal{L}_{localize}$, and CodeT5-R trained with $\mathcal{L}_{repair}$. We set the maximum source and target sequence lengths to 512. All experiments are performed on NVIDIA A100 GPUs with 40 GB memory. 

\subsection{Function-Level Bug Detection}
\paragraph{Metrics}
For this task, we use two metrics: the F1 score and the False Positive Rate (FPR). F1 is the standard metric for this type of task because it is a binary classification problem (buggy or not). The FPR, on the other hand, is critical for determining a bug localization system's usability in a real-world scenario. A good bug detection system should produce as few false positives as possible~\cite{allamanis2021self, vasic2019neural}.  FPR is calculated as the ratio between the number of non-buggy functions wrongly categorized as buggy (false positives) and the total number of actual non-buggy functions. 

\paragraph{Baselines}
The function-level bug detection can seen as the code classification task, i.e, assign a label to a given code snippet.  We choose Tree-based CNN~\cite{mou2016convolutional}, a well-known method for code classification as a baseline. We also including some others SOTA pretrained language models of code, which are CodeBERT~\cite{feng-etal-2020-codebert}, GraphCodeBERT~\cite{guo2020graphcodebert}, and PLBART~\cite{ahmad2021unified}. We used their public checkpoints and fine-tune them for this task. We also include SpotBugs~\footnote{\url{https://github.com/spotbugs/spotbugs}}, a widely used static analysis-based baseline~\cite{karampatsis2020often, habib2018many} for bug detection task. For CodeT5, we use 3 baselines: CodeT5-L, CodeT5-D and CodeT5-DLR. CodeT5-R is not trained for bug detection but its output can also be used to detect bug~\footnote{When perform training, we feed the model with both negative and positive samples so the PR module is able to decide to generate fixed code or not}.

\begin{table}[t]
\centering
\caption{Performance of function-level bug detection. $\uparrow$: the higher the better, $\downarrow$: the lower the better.}
\label{tab:detect}
\resizebox{0.5\textwidth}{!} {
\begin{tabular}{l|cc|cc}
	\hline
	\multirow{2}{*}{\textbf{Model}} & \multicolumn{2}{c|}{SL-Java} & \multicolumn{2}{c}{ML-Python} \\ \cline{2-5} 
	& F1~$\uparrow$           & FPR~$\downarrow$           & F1~$\uparrow$            & FPR~$\downarrow$           \\ \hline
	SpotBugs                      & 4.6        & 89.29        & -       & -       \\
	TBCNN                      & 45.49        & 38.03        & 51.24         & 48.20         \\
	CodeBERT                   & 55.67        & 40.02        & 50.24         & 50.32         \\
	GraphCodeBERT              & 56.44        & 38.99        & 49.30         & 53.20         \\
	PLBART                     & 59.01        & 35.21        & 52.33         & 51.24         \\ \hline
	CodeT5-R      & 50.94        & 40.29        & 52.93         & 46.45         \\
	CodeT5-D     & 59.28        & 34.32        & 52.93         & 46.45         \\
	CodeT5-DLR    & \textbf{63.46}        & \textbf{31.24}        & \textbf{54.83}         & \textbf{43.21 }        \\ \hline
\end{tabular}
}
\end{table}
\paragraph{Results}

Table~\ref{tab:detect} shows the results of function-level bug detection task. Our model fine-tuned with all 3 objectives together achieve the best performance in terms of both F1 score and FPR, while our CodeT5-D with  the  only function-level bug detection objective still yields better results than the baselines. The SpotBugs baseline achieves only 4.6 in F1 score, which is consistent with the performance reported in~\citet{habib2018many} for static analysis-based bug detector.

\subsection{Line-Level Bug Localization}
\begin{table*}[t]
	\centering
	\caption{Performance of line-level bug localization. $\uparrow$: the higher the better and $\downarrow$: the lower the better.}
	\label{tab:localize}
	\resizebox{1.0\textwidth}{!} {
		\begin{tabular}{l|cccc|cccc}
			\hline
			\multirow{2}{*}{\textbf{Model}} & \multicolumn{4}{c|}{SL-Java}   & \multicolumn{4}{c}{ML-Python} \\ \cline{2-9} 
			& MRR@1~$\uparrow$      & MRR@5~$\uparrow$      & FPR@1~$\downarrow$     & FPR@5~$\downarrow$ & MAP@1 ~$\uparrow$     & MAP@5~$\uparrow$      & FPR@1~$\downarrow$ & FPR@5~$\downarrow$ \\ \hline
			DeepLineDP                 & 14.05 & 15.33 & 25.30 & 41.20 & 9.35  & 11.98 & 25.60 & 73.21 \\
			LineVul                    & 15.34 & 16.79 & 23.59 & 39.45 & 10.46 & 14.24 & 27.46 & 67.22 \\
			CodeBERT                   & 21.66 & 29.69 & 17.20 & 21.57 & 18.67 & 26.30 & 19.92 & 58.90 \\
			GraphCodeBERT              & 20.35 & 28.45 & 18.67 & 20.45 & 22.59 & 28.92 & 19.20 & 53.24 \\
			PLBART                     & 23.02 & 30.98 & 12.93 & 14.67 & 23.22 & 30.56 & 13.98 & 46.22 \\ \hline
			CodeT5-L      & 24.40 & 32.01 & 7.33  & 10.39 & 24.59 & 32.40 & 13.67 & 42.44 \\
			CodeT5-DLR    & \textbf{26.78} & \textbf{34.67} & \textbf{3.04}  & \textbf{8.05}  & \textbf{26.98} & \textbf{33.75} & \textbf{9.84}  & \textbf{38.46} \\ 
			CodeT5-DLR-new    & \textbf{27.67} & \textbf{38.38} & \textbf{3.01}  & \textbf{7.23}  & \textbf{-} & \textbf{-} & \textbf{-}  & \textbf{-} \\ \hline
		\end{tabular}
x	}
\end{table*}

\paragraph{Metrics}
We use 3 metrics for this task, which are Mean Reciprocal Rank (MRR), Mean Average Precision (MAP), and False Positive Rate (FPR).
In reality, we do not know how many lines of code is buggy, so we retrieve top-k lines with highest scores and measure if the ground-truth buggy lines(s) belong to these top-k lines. As such, we can formulate this problem as an information retrieval problem, with the goal of returning a ranked list of relevant lines to a query (the query is to retrieve all of the buggy lines among the lines). For this reason, we use the well-known metrics of MRR and MAP to evaluate for this buggy line retrieval task. In our evaluation settings, each of these metrics will be appropriate for a different datasets. For the SL-Java, because there is only one buggy line in the ground truth, the MRR is appropriate for evaluating the performance of this dataset. On the other hand, each of sample in ML-Python contains numerous buggy lines in the ground truth, the MAP is better suited for ML-Python. With this, MRR and MAP are computed with respect to $k$, resulting in MRR@k and MAP@k, where $k$ is number of lines retrieved for evaluation. We choose $k = 1$ and $k = 5$ for our evaluation. 

%\paragraph{Mean Reciprocal Rank} : The mean reciprocal rank is the average of the reciprocal ranks of results of a set of
%queries Q. The reciprocal rank of a query is the multiplicative inverse of the rank of the first correct answer. MRR can be calculated as follows:
%\begin{equation}
%	\small
%	MRR = \frac{1}{\mid Q\mid}\sum_{i=1}^{\mid Q\mid}\frac{1}{\mid rank_{i}\mid}
%\end{equation}
%
%\paragraph{Mean Average Precision} : Mean average precision (MAP) for a set of queries is the mean of the average precision scores for each query. MAP provides a single-figure measure of quality of information retrieval, when a query may have multiple relevant documents. The Average Precision (AveP) of a single query is the average of the precision values obtained for the query. MAP and AveP can be calculated as follows:
%\begin{equation}
%	\small
%	AveP = \frac{1}{\mid Q\mid}\sum_{i=1}^{\mid M\mid}\frac{1}{\mid rank_{i}\mid};
%	\\
%	MAP = \frac{\sum_{q=1}^{Q}AveP(q)}{Q}
%\end{equation}

In addition to MRR and MAP, we use False Positive Rate (FPR) to evaluate. Given a code snippet and retrieved buggy lines, the FPR in this case is calculated as the ratio between the number of non-buggy lines wrongly categorized as buggy (false positives) and the total number of actual non-buggy lines. We also compute FPR with respected to top-k lines retrieved, similar to MRR and MAP. 
\paragraph{Baselines}
For this task, we also chose baselines that are similar to the Function-level bug localization task, which are CodeBERT~\cite{feng-etal-2020-codebert}, GraphCodeBERT~\cite{guo2020graphcodebert}, and PLBART~\cite{ahmad2021unified}. In addition, we include 2 additional baselines that have been used to detect vulnerability in software engineering, which are DeepLineDP~\cite{pornprasit2022deeplinedp} and LineVul~\cite{fu2022linevul}. They work by simply performing prediction at the function level, then using attention scores from the backbone neural architecture to retrieve the line scores to predict vulnerability at line level. DeepLineDP is based on the Hierrarchical Attention Network~\cite{yang2016hierarchical}, which divides the source code into three layers: function, line, and token, with each level processed by a BiGRU neural network. LineVul is based on a vanilla Transformer~\cite{vaswani2017attention}, and its scores are calculated by averaging the token scores from the multi head attention layer. DeepLineDP and LineVul has not been used for bug localization before, but we try our best to adapt their software artifacts~\footnote{\url{https://github.com/awsm-research/DeepLineDP}}~\footnote{\url{https://github.com/awsm-research/LineVul}} into our use case.

\paragraph{Results}

The results of line-level bug localization task are shown in Table~\ref{tab:localize}.  When fine-tuning on all objectives, our model outperforms all baselines in terms of all metrics. The FPR is low when we only aim to detect one line of buggy code. When we increase k to 5, the FPR increases. However, the MRR and MAP are also better for either single-line bug detection (SL-Java) or multi-line bug detection (ML-Python) with k = 5. It means that as we broaden the scope of buggy line retrieval, the number of correctly detected bugs increases, but the model produces more false alarms. When performing bug localization at the line level, this is a trade off that must be made.

\begin{table}[ht]
	\centering
	\caption{Performance of Program Repair task. }
	\label{tab:repair}
	\resizebox{0.5\textwidth}{!} {
	\begin{tabular}{l|cc|cc}
		\hline
		\multirow{2}{*}{\textbf{Model}} & \multicolumn{2}{c|}{SL-Java} & \multicolumn{2}{c}{ML-Python} \\ \cline{2-5} 
		& EM           & BLEU         & EM            & BLEU          \\ \hline
		CodeBERT                   & 3.66         & 34.21        & 3.40          & 29.40         \\
		GraphCodeBERT              & 3.35         & 35.29        & 3.31          & 30.59         \\
		PLBART                     & 6.02         & 40.12        & 5.39          & 33.92         \\ \hline
		CodeT5-R      & 7.30         & 40.20        & 6.01          & 35.54         \\
		CodeT5-DLR    & \textbf{10.30}        & \textbf{43.42}        & \textbf{6.30}          & \textbf{38.44}         \\ \hline
	\end{tabular}
	}
\end{table}
\subsection{Program Repair}
\paragraph{Metrics}
We use 2 metrics for program repair: Exact Match (EM) and BLEU. For EM, if the generated program exactly matches the ground truth correct program, then EM=1, otherwise EM=0. BLEU score is a standard metric that is usually used to measure translation-based tasks.

\paragraph{Baselines}
We also use CodeBERT~\cite{feng-etal-2020-codebert}, GraphCodeBERT~\cite{guo2020graphcodebert}, and PLBART~\cite{ahmad2021unified} for the program repair task. We fine-tune these pretrained models with the $\mathcal{L}_{repair}$ objective to generate the fixed code from buggy code.

\paragraph{Results}

Table~\ref{tab:repair} shows the results of program repair task. Our model when fine-tuning on all 3 objectives achieves the best performance among the baselines with significant margins, both in terms of EM and BLEU. In overall, EM for ML-Python is lower than EM for SL-Java, it is because that it is more challenging to generate fixed code given that there are multi buggy lines in the buggy code.

\subsection{Qualitative Analysis}
\subsubsection{End-to-End Bug Detection and Repair}
\label{sec:end_to_end}
\begin{figure}[t]
	\centering
	\includegraphics[width=0.5\textwidth]{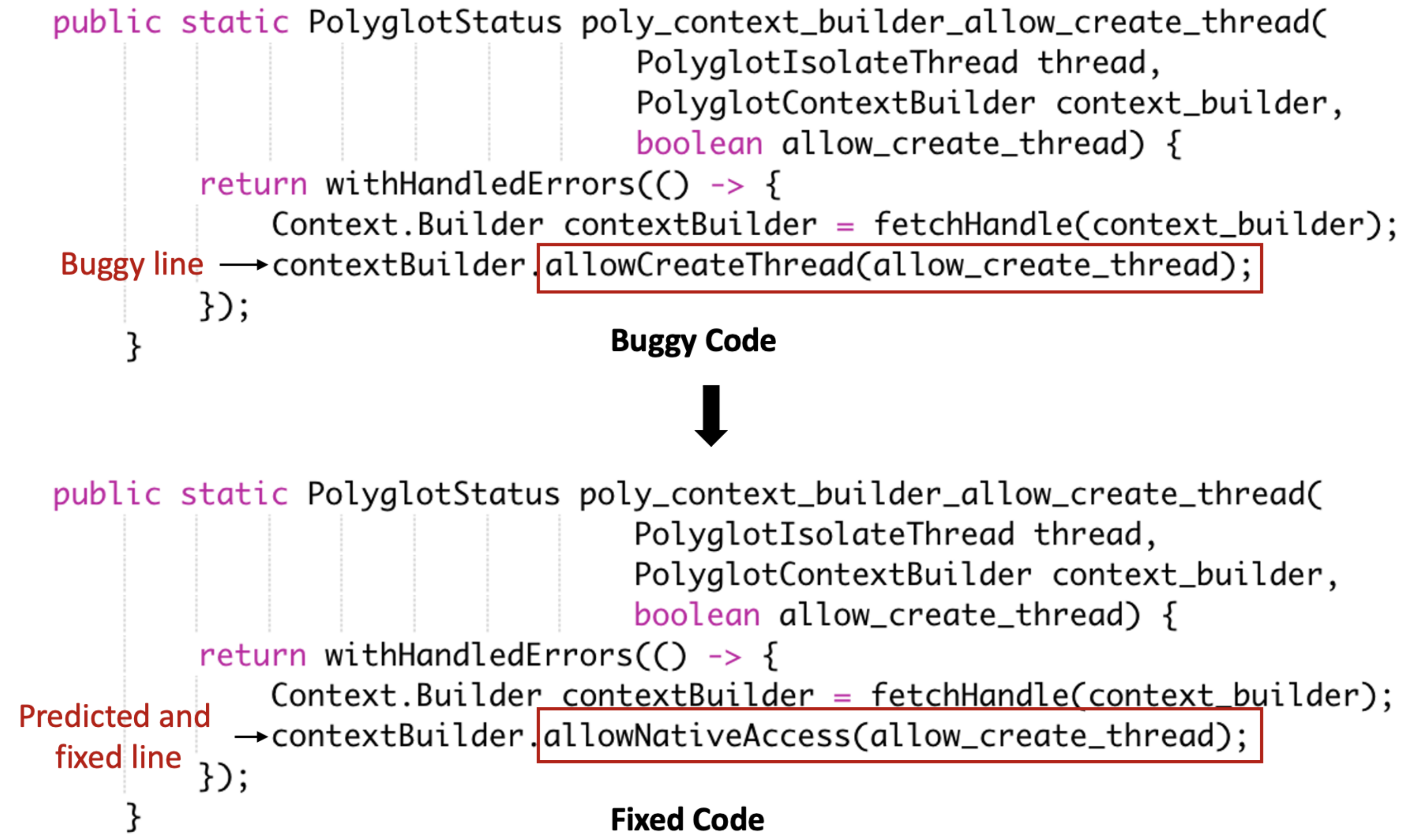}
	\caption{A \textit{CHANGE\_CALLER\_IN\_FUNCTION} bug that our Code-DLR can successfully detect and repair.}
	\label{fig:sample_1}
\end{figure}

We have shown that our model performs the best among all of the baselines for 3 tasks: function-level bug detection, line-level bug localization and program repair. However, since these 3 tasks are evaluate individually, they still do not reflect the full capability of our model in a unified manner, i.e., both detect bugs and suggest fixes. It also reflects how the developer debugs program in their daily work. We perform additional experiments to illustrate these steps in order. 
First, we use the function-level bug detection module to predict a set of buggy functions from the test set, regardless of whether they are buggy or not. Then, using the detected samples, we use the line-level bug localization module to identify buggy lines within these detected samples (not all samples in the test set). We then use the program repair module to suggest fixes for these samples as well. Figure~\ref{fig:sample_1} shows a bug example that our model can detect, localize and repair. Note that this is a real example from an open source project~\footnote{\url{https://github.com/oracle/graal/commit/9059770c00748fac01a75bac6f30d074783d2e69}} with 17K stars on Github. This is a real bug-fix commit with the commit message "\textit{Minor fix in polyglot native API}". First, our function-level bug detection model can detect that this is a buggy line. Second, the line-level bug localization model ranks the line \textit{contextBuilder.allowNativeAccess(allow\_create\_thread);} as the top-1 line that is buggy. Finally, the program repair module translates the whole buggy function to the fixed function, and the buggy part  "\texttt{allowCreateThread}" is translated to the correct version " \texttt{allowNativeAccess}". This fix could be referred to the \textit{CHANGE\_CALLER\_IN\_FUNCTION} pattern, in which the invoked function of an object is changed.

\begin{figure}[t]
	\centering
	\includegraphics[width=0.5\textwidth]{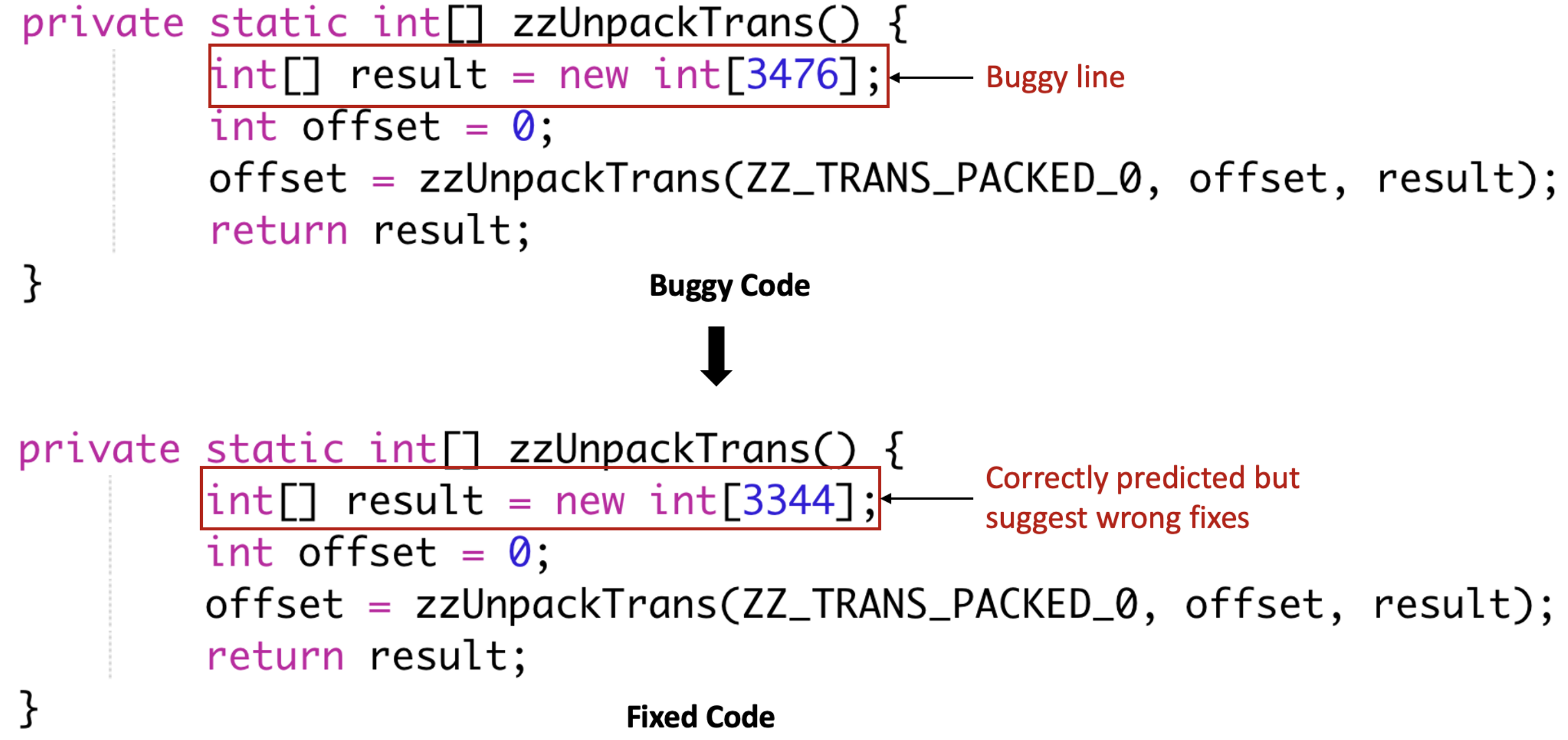}
	\caption{A \textit{CHANGE\_NUMERAL} bug that our Code-DLR can successfully detect but suggest wrong fixes.}
	\label{fig:sample_3}
\end{figure}

Figure~\ref{fig:sample_3} shows another example. This is yet another bug found in a real-world project~\footnote{\url{https://github.com/apache/druid/commit/7ebe053ac1abce6e3b218beaee801ebbc6da2ecb}}. Our DLR process can also identify the correct buggy line but fails to recommend correct fixes. However, this is due to the fact that the fix is for the pattern \textit{CHANGE\_NUMERAL}, which is very challenging to know the exact numeral to replace (3476 to 3344).  An enhancement to our technique is the ability to suggest fixes for the missing whole of a broken line, similar to ~\citet{guo2021learning} for code completion. We leave this as a part of future investigation.
%Due to the page constraints, we include more success and failed samples in the Appendix.
%~\footnote{https://github.com/jindrapetrik/jpexs-decompiler/commit/6df9ea4e76fed6036e50a64410da828080fd48de}
%

%\begin{figure}[t]
%	\centering
%	\includegraphics[width=0.5\textwidth]{figures/sample_2.pdf}
%	\caption{A \textit{CHANGE\_UNARY\_OPERATOR} bug that our Code-DLR can successfully detect and repair.}
%	\label{fig:sample_2}
%\end{figure}

\begin{figure*}[t]
	\centering
	\hspace*{-2.1cm} 
	\includegraphics[width=1.1\textwidth]{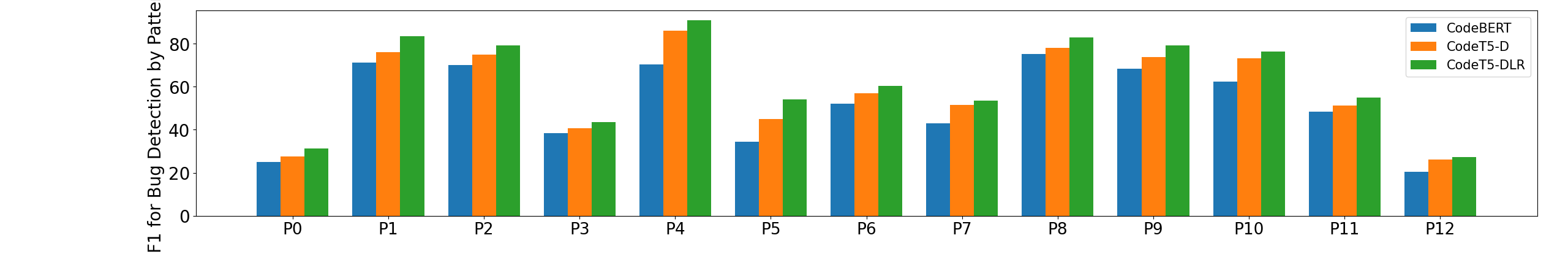}
		\vspace{-0.1in}
	\caption{Analysis of bug detection in F1 score, broken down by 14 bug patterns.}
	\label{fig:bug_patterns}
	\vspace{-0.1in}
\end{figure*}

\subsubsection{Analysis of Detected Bug Patterns}
The SL-Java dataset contains bug pattern information, analyzing and understanding how different bug patterns can be detected and fixed is critical for future debugging system improvement. We break down the performance by bug-pattern and computer the percentage of successfully detected bugs for each of the patterns in terms of F1 score based on the results of function-level bug detection in Table~\ref{tab:detect}
We use 3 baselines for this analysis, which are: CodeBert, CodeT5-L. and CodeT5-DLR.
Figure~\ref{fig:bug_patterns} illustrate such results. CodeT5-L performs better than CodeT5-DLR on an average of 4.1\%, and is better than CodeBert on an average of 10.2\% in terms of F1.
For some patterns, such as P1(change operand), P2 (change identifier) and P5 (change caller in fuction), CodeT5-DLR perform much better than CoderBert (>14\% on average)~\footnote{Due to page number constraints, we were unable to display the pattern name on the bar chart; instead, we abstracted the pattern into some index, such as P0, P1, and so on. Readers are encouraged to view the explanations for each of the patterns in our supplementary materials.}.

\subsection{Analysis of End-to-End Bug Detection and Repair}
In addition to the samples shown in Section~\ref{sec:end_to_end},  we also want to see how the other models compare to ours in this end-to-end process. However, because we discovered no baseline in the literature that performs this unify task in one model, we were unable to compare with the other baselines. As such, we compare this to the models that are trained individually for each task and perform the same debugging steps as described.
Table~\ref{tab:unify} shows that CodeT5-DLR continues to outperform its variants. Note that the performance of CodeT5-DLR this Table for line-level bug localization (BL) and program repair (PR) are different than the same model for the same tasks in Table~\ref{tab:localize} and Table~\ref{tab:repair}.
\begin{table}[ht]
	\centering
	\caption{Performance of Unify Debugging Procedure. CodeT5-L performs worse than CodeT5-DLR for the line-level bug localization (BL) task in term of MRR@5. CodeT5-R also performs worse than CodeT5-DLR for the program repair (PR) task in term of BLEU.}

	\resizebox{0.5\textwidth}{!} {
	\begin{tabular}{l|cc|cc}
    \hline
    \multirow{2}{*}{\textbf{Model}}                & \multicolumn{2}{c|}{SL-Java} & \multicolumn{2}{c}{ML-Python}      \\ \cline{2-5} 
                                              & BL     & PR    & BL & PR              \\ \hline
  CodeT5-L  & 28.58         & -           & 24.59      & -\\
    CodeT5-R                   & -             & 43.20       & -          & 38.50                 \\ \hline
    CodeT5-DLR                & 33.93         & 46.93       & 28.49      & 41.21                 \\ \hline
    \end{tabular}
	}
	\label{tab:unify}
\end{table}

\section{Discussion}
\label{sec:diss}
In this section, we discuss the design choice of the model architecture, as well as some advantages and disadvantages in comparison to other techniques. A recently developed line of work also employs a joint objective function for bug localization and repair~\cite{allamanis2021self, vinyals2015pointer}. However, the task they are attempting is not the same as ours. They aim to detect bugs at token level, where the token is considered as a missing slot in the code representation and the goal is to fill in such missing slot. A pointer network~\cite{vinyals2015pointer} is leveraged to predict whether a location is buggy or not. In fact, we can also design a pointer network to detect whether a line is buggy or not in the line-level bug localization objective (instead of the current ranking method). However, because we want to detect both buggy and non-buggy code, this design is ineffective in our case. Their method is based on the assumption that the buggy code is given, the goal is to locate the bug. We may end up feeding the pointer net a lot of non-buggy lines, which usually have a much larger number than buggy lines. Because of the imbalance between buggy and non-buggy lines, the pointer net design does not work well in our case (confirmed with experiments).

In addition, their models are impractical in real-world use case due to high FPR(~98\% in \citet{allamanis2021self}). Also, the evaluation process of these techniques is mostly based on synthetic datasets, i.e., the bugs are generated using some heuristics, making the evaluation results unrealistic~\cite{he2022distribution}. By contrast, our work does not rely on synthetic bug data, but rather on real-world datasets from Github projects, making the results more useful in practice. In addition, the granularity of bugs we are targeting is different so that our technique cannot be directly compared to theirs.

%Our intention to design these 3 objectives are that 2 objectives will assist the remaining to make the specific task for that objective becomes better than when training invididually. However, this does not come with the consistent signal across the 3 modules, e.g., $\mathcal{L}_{detection}$ decide if . In the ideal case, if $\mathcal{L}_{detection}$ decides if a given function is not buggy, $\mathcal{L}_{repair}$ should not generate any fixes for such function. However, our result shows that there are many cases that they are not consistent, this requires further investigation on improving the model designs

%A potential way to address this issue is that we can train a Program Repair model only on negative bug-fix pairs only, this mean that the model will always suggest to fix something from the input program (regardless if it is buggy or not). This will make sure that if a program
\section{Related Work} \label{sec:related_work}
\paragraph{Pretrained Language Models for Code}
Recently, language models in natural language processing has been applied  to model source code~\cite{feng-etal-2020-codebert,wang2021codet5, guo2020graphcodebert, ahmad2021unified, bui2021self, elnaggar2021codetrans,peng2021could, kanade2020learning}.  CodeBERT~\cite{feng-etal-2020-codebert} pretrain a model of code on multiple programming languages by adapting a Roberta model~\cite{liu2019roberta}.
CuBERT~\cite{kanade2020learning} pretrains a BERT model for code using a large dataset of curated Python files. In general, most of the techniques treat code similar to texts and adapt the same pretraining strategies as for natural language. 
Some techniques, such as CodeT5~\cite{wang2021codet5}, encode source code features such as identifier information, data flow, and function name, among others, to pretrain code models, which may result in better overall performance.

\paragraph{Neural-based Bug Localization and Program Repair}
Bug localization and program repairs have received a lot of attention in terms of combining language models with traditional static analysis-based methods to improve performance. A recent trend is to generate synthetic simple bugs by rewriting rules into programs and then use self-supervised learning to train jointly models for bug localization and repair~\cite{allamanis2021self,yasunaga2021break,yasunaga2020graph,vasic2019neural}. However, these techniques are almost impractical for real-world use case since they only target \textit{simple bugs} and the models are trained mostly on synthetic data. There are also many recent neural-based techniques that target only program repairs ~\cite{lutellier2020coconut,zhu2021syntax,jiang2021cure,chen2019sequencer,li2020dlfix, tufano2018empirical}. In contrast, our CodeT5-DLR aims to combine the strengths of each of these techniques in order to fine-tune a foundation model for jointly localizing bugs and repairing programs at a reasonable code granularity (function and line level).

%These techniques work by learning change patterns from code changes and assuming that the buggy parts of the code are provided. As a result, they lack the ability to determine whether the inputs are truly buggy and continue to generate patches.

\section{Conclusion} \label{sec: conclusion}
We proposed a novel detect-localize-repair framework for jointly detecting bugs, localizing bugs and suggesting program repairs. Our model is built on the CodeT5 foundation model and is fine-tuned jointly to achieve three debugging-related objectives: function-level bug detection, line-level bug localization, and program repair. These three objectives are based on how software developers locate bugs and repair programs in their daily work. Our evaluation results show that training these 3 objectives together yields better results than fine-tuning on each objective individually. Furthermore, we also contribute to provide two new datasets that can be used for evaluating both bug localization and program repair tasks. Our datasets differ from existing datasets in that we provide the exact line that is buggy, as well as the before and after versions of a code snippet. We will make our datasets publicly available to facilitate research on this topic.

\newpage
\section{Limitations}
\label{sec:limit}
There are still a few drawbacks of our technique that need further investigations. Our intention in designing these three objectives is that each of the two will assist the third in making the specific task for that objective better than when training individually. However, this does not provide a consistent signal across the three modules. For example, while the function-level bug detection module indicates that a function is not buggy, the program repair module continues to generate fixes. This inconsistency has been reflected in Table~\ref{tab:detect}. A future research direction would be to design the training pipeline step by step, with the program repair module only providing repairs based on the bug detection module's signal. Second, we only use within-function information for the three tasks. A function in a program, on the other hand, is usually linked to other parts of the program or other functions within the same file. This information is commonly referred to as contexts, and it should be used as additional information when localizing bugs. We leave these investigations for the future work.

\bibliographystyle{acl_natbib}
\bibliography{references}

\begin{thebibliography}{45}
\expandafter\ifx\csname natexlab\endcsname\relax\def\natexlab#1{#1}\fi

\bibitem[{Abreu et~al.(2007)Abreu, Zoeteweij, and
  Van~Gemund}]{abreu2007accuracy}
Rui Abreu, Peter Zoeteweij, and Arjan~JC Van~Gemund. 2007.
\newblock On the accuracy of spectrum-based fault localization.
\newblock In \emph{Testing: Academic and industrial conference practice and
  research techniques-MUTATION (TAICPART-MUTATION 2007)}, pages 89--98. IEEE.

\bibitem[{Abreu et~al.(2009)Abreu, Zoeteweij, and
  Van~Gemund}]{abreu2009spectrum}
Rui Abreu, Peter Zoeteweij, and Arjan~JC Van~Gemund. 2009.
\newblock Spectrum-based multiple fault localization.
\newblock In \emph{2009 IEEE/ACM International Conference on Automated Software
  Engineering}, pages 88--99. IEEE.

\bibitem[{Ahmad et~al.(2021)Ahmad, Chakraborty, Ray, and
  Chang}]{ahmad2021unified}
Wasi~Uddin Ahmad, Saikat Chakraborty, Baishakhi Ray, and Kai{-}Wei Chang. 2021.
\newblock {Unified Pre-training for Program Understanding and Generation}.
\newblock In \emph{Proceedings of the 2021 Conference of the North American
  Chapter of the Association for Computational Linguistics: Human Language
  Technologies, {NAACL-HLT} 2021, Online, June 6-11, 2021}, pages 2655--2668.
  Association for Computational Linguistics.

\bibitem[{Allamanis et~al.(2021)Allamanis, Jackson-Flux, and
  Brockschmidt}]{allamanis2021self}
Miltiadis Allamanis, Henry Jackson-Flux, and Marc Brockschmidt. 2021.
\newblock Self-supervised bug detection and repair.
\newblock \emph{Advances in Neural Information Processing Systems},
  34:27865--27876.

\bibitem[{Bui et~al.(2021)Bui, Yu, and Jiang}]{bui2021self}
Nghi~DQ Bui, Yijun Yu, and Lingxiao Jiang. 2021.
\newblock Self-supervised contrastive learning for code retrieval and
  summarization via semantic-preserving transformations.
\newblock In \emph{Proceedings of the 44th International ACM SIGIR Conference
  on Research and Development in Information Retrieval}, pages 511--521.

\bibitem[{Chen et~al.(2019)Chen, Kommrusch, Tufano, Pouchet, Poshyvanyk, and
  Monperrus}]{chen2019sequencer}
Zimin Chen, Steve Kommrusch, Michele Tufano, Louis-No{\"e}l Pouchet, Denys
  Poshyvanyk, and Martin Monperrus. 2019.
\newblock Sequencer: Sequence-to-sequence learning for end-to-end program
  repair.
\newblock \emph{IEEE Transactions on Software Engineering}, 47(9):1943--1959.

\bibitem[{Ding et~al.(2020)Ding, Ray, Devanbu, and
  Hellendoorn}]{ding2020patching}
Yangruibo Ding, Baishakhi Ray, Premkumar Devanbu, and Vincent~J Hellendoorn.
  2020.
\newblock Patching as translation: the data and the metaphor.
\newblock In \emph{2020 35th IEEE/ACM International Conference on Automated
  Software Engineering (ASE)}, pages 275--286. IEEE.

\bibitem[{Elnaggar et~al.(2021)Elnaggar, Ding, Jones, Gibbs, Feher, Angerer,
  Severini, Matthes, and Rost}]{elnaggar2021codetrans}
Ahmed Elnaggar, Wei Ding, Llion Jones, Tom Gibbs, Tamas Feher, Christoph
  Angerer, Silvia Severini, Florian Matthes, and Burkhard Rost. 2021.
\newblock Codetrans: Towards cracking the language of silicon's code through
  self-supervised deep learning and high performance computing.
\newblock \emph{arXiv preprint arXiv:2104.02443}.

\bibitem[{Feng et~al.(2020)Feng, Guo, Tang, Duan, Feng, Gong, Shou, Qin, Liu,
  Jiang, and Zhou}]{feng-etal-2020-codebert}
Zhangyin Feng, Daya Guo, Duyu Tang, Nan Duan, Xiaocheng Feng, Ming Gong, Linjun
  Shou, Bing Qin, Ting Liu, Daxin Jiang, and Ming Zhou. 2020.
\newblock \href {https://doi.org/10.18653/v1/2020.findings-emnlp.139}
  {{C}ode{BERT}: A pre-trained model for programming and natural languages}.
\newblock In \emph{Findings of the Association for Computational Linguistics:
  EMNLP 2020}, pages 1536--1547, Online. Association for Computational
  Linguistics.

\bibitem[{Fu and Tantithamthavorn(2022)}]{fu2022linevul}
Michael Fu and Chakkrit Tantithamthavorn. 2022.
\newblock Linevul: A transformer-based line-level vulnerability prediction.

\bibitem[{Guo et~al.(2020)Guo, Ren, Lu, Feng, Tang, Liu, Zhou, Duan,
  Svyatkovskiy, Fu et~al.}]{guo2020graphcodebert}
Daya Guo, Shuo Ren, Shuai Lu, Zhangyin Feng, Duyu Tang, Shujie Liu, Long Zhou,
  Nan Duan, Alexey Svyatkovskiy, Shengyu Fu, et~al. 2020.
\newblock Graphcodebert: Pre-training code representations with data flow.
\newblock \emph{arXiv preprint arXiv:2009.08366}.

\bibitem[{Guo et~al.(2021)Guo, Svyatkovskiy, Yin, Duan, Brockschmidt, and
  Allamanis}]{guo2021learning}
Daya Guo, Alexey Svyatkovskiy, Jian Yin, Nan Duan, Marc Brockschmidt, and
  Miltiadis Allamanis. 2021.
\newblock Learning to complete code with sketches.
\newblock In \emph{International Conference on Learning Representations}.

\bibitem[{Habib and Pradel(2018)}]{habib2018many}
Andrew Habib and Michael Pradel. 2018.
\newblock How many of all bugs do we find? a study of static bug detectors.
\newblock In \emph{2018 33rd IEEE/ACM International Conference on Automated
  Software Engineering (ASE)}, pages 317--328. IEEE.

\bibitem[{He et~al.(2022)He, Beurer-Kellner, and Vechev}]{he2022distribution}
Jingxuan He, Luca Beurer-Kellner, and Martin Vechev. 2022.
\newblock On distribution shift in learning-based bug detectors.
\newblock \emph{arXiv preprint arXiv:2204.10049}.

\bibitem[{Hindle et~al.(2016)Hindle, Barr, Gabel, Su, and
  Devanbu}]{hindle2016naturalness}
Abram Hindle, Earl~T Barr, Mark Gabel, Zhendong Su, and Premkumar Devanbu.
  2016.
\newblock On the naturalness of software.
\newblock \emph{Communications of the ACM}, 59(5):122--131.

\bibitem[{Jia and Harman(2010)}]{jia2010analysis}
Yue Jia and Mark Harman. 2010.
\newblock An analysis and survey of the development of mutation testing.
\newblock \emph{IEEE transactions on software engineering}, 37(5):649--678.

\bibitem[{Jiang et~al.(2021)Jiang, Lutellier, and Tan}]{jiang2021cure}
Nan Jiang, Thibaud Lutellier, and Lin Tan. 2021.
\newblock Cure: Code-aware neural machine translation for automatic program
  repair.
\newblock In \emph{2021 IEEE/ACM 43rd International Conference on Software
  Engineering (ICSE)}, pages 1161--1173. IEEE.

\bibitem[{Kanade et~al.(2020)Kanade, Maniatis, Balakrishnan, and
  Shi}]{kanade2020learning}
Aditya Kanade, Petros Maniatis, Gogul Balakrishnan, and Kensen Shi. 2020.
\newblock Learning and evaluating contextual embedding of source code.
\newblock In \emph{International Conference on Machine Learning}, pages
  5110--5121. PMLR.

\bibitem[{Karampatsis and Sutton(2020)}]{karampatsis2020often}
Rafael-Michael Karampatsis and Charles Sutton. 2020.
\newblock How often do single-statement bugs occur? the manysstubs4j dataset.
\newblock In \emph{Proceedings of the 17th International Conference on Mining
  Software Repositories}, pages 573--577.

\bibitem[{Le et~al.(2013)Le, Thung, and Lo}]{le2013theory}
Tien-Duy~B Le, Ferdian Thung, and David Lo. 2013.
\newblock Theory and practice, do they match? a case with spectrum-based fault
  localization.
\newblock In \emph{2013 IEEE International Conference on Software Maintenance},
  pages 380--383. IEEE.

\bibitem[{Li et~al.(2020)Li, Wang, and Nguyen}]{li2020dlfix}
Yi~Li, Shaohua Wang, and Tien~N Nguyen. 2020.
\newblock Dlfix: Context-based code transformation learning for automated
  program repair.
\newblock In \emph{Proceedings of the ACM/IEEE 42nd International Conference on
  Software Engineering}, pages 602--614.

\bibitem[{Liu et~al.(2019)Liu, Ott, Goyal, Du, Joshi, Chen, Levy, Lewis,
  Zettlemoyer, and Stoyanov}]{liu2019roberta}
Yinhan Liu, Myle Ott, Naman Goyal, Jingfei Du, Mandar Joshi, Danqi Chen, Omer
  Levy, Mike Lewis, Luke Zettlemoyer, and Veselin Stoyanov. 2019.
\newblock Roberta: A robustly optimized bert pretraining approach.
\newblock \emph{arXiv preprint arXiv:1907.11692}.

\bibitem[{Lu et~al.(2021)Lu, Guo, Ren, Huang, Svyatkovskiy, Blanco, Clement,
  Drain, Jiang, Tang, Li, Zhou, Shou, Zhou, Tufano, Gong, Zhou, Duan,
  Sundaresan, Deng, Fu, and LIU}]{lu2021codexglue}
Shuai Lu, Daya Guo, Shuo Ren, Junjie Huang, Alexey Svyatkovskiy, Ambrosio
  Blanco, Colin Clement, Dawn Drain, Daxin Jiang, Duyu Tang, Ge~Li, Lidong
  Zhou, Linjun Shou, Long Zhou, Michele Tufano, Ming Gong, Ming Zhou, Nan Duan,
  Neel Sundaresan, Shao~Kun Deng, Shengyu Fu, and Shujie LIU. 2021.
\newblock \href {https://openreview.net/forum?id=6lE4dQXaUcb} {Code{XGLUE}: A
  machine learning benchmark dataset for code understanding and generation}.
\newblock In \emph{Thirty-fifth Conference on Neural Information Processing
  Systems Datasets and Benchmarks Track (Round 1)}.

\bibitem[{Lutellier et~al.(2020)Lutellier, Pham, Pang, Li, Wei, and
  Tan}]{lutellier2020coconut}
Thibaud Lutellier, Hung~Viet Pham, Lawrence Pang, Yitong Li, Moshi Wei, and Lin
  Tan. 2020.
\newblock Coconut: combining context-aware neural translation models using
  ensemble for program repair.
\newblock In \emph{Proceedings of the 29th ACM SIGSOFT international symposium
  on software testing and analysis}, pages 101--114.

\bibitem[{Mashhadi and Hemmati(2021)}]{mashhadi2021applying}
Ehsan Mashhadi and Hadi Hemmati. 2021.
\newblock Applying codebert for automated program repair of java simple bugs.
\newblock In \emph{2021 IEEE/ACM 18th International Conference on Mining
  Software Repositories (MSR)}, pages 505--509. IEEE.

\bibitem[{Mou et~al.(2016)Mou, Li, Zhang, Wang, and Jin}]{mou2016convolutional}
Lili Mou, Ge~Li, Lu~Zhang, Tao Wang, and Zhi Jin. 2016.
\newblock Convolutional neural networks over tree structures for programming
  language processing.
\newblock In \emph{Thirtieth AAAI conference on artificial intelligence}.

\bibitem[{Papadakis and Le~Traon(2015)}]{papadakis2015metallaxis}
Mike Papadakis and Yves Le~Traon. 2015.
\newblock Metallaxis-fl: mutation-based fault localization.
\newblock \emph{Software Testing, Verification and Reliability},
  25(5-7):605--628.

\bibitem[{Peng et~al.(2021)Peng, Zheng, Li, Ke, He, and Liu}]{peng2021could}
Dinglan Peng, Shuxin Zheng, Yatao Li, Guolin Ke, Di~He, and Tie-Yan Liu. 2021.
\newblock How could neural networks understand programs?
\newblock In \emph{International Conference on Machine Learning}, pages
  8476--8486. PMLR.

\bibitem[{Pornprasit and Tantithamthavorn(2022)}]{pornprasit2022deeplinedp}
Chanathip Pornprasit and Chakkrit Tantithamthavorn. 2022.
\newblock Deeplinedp: Towards a deep learning approach for line-level defect
  prediction.
\newblock \emph{IEEE Transactions on Software Engineering}.

\bibitem[{Ray et~al.(2016)Ray, Hellendoorn, Godhane, Tu, Bacchelli, and
  Devanbu}]{ray2016naturalness}
Baishakhi Ray, Vincent Hellendoorn, Saheel Godhane, Zhaopeng Tu, Alberto
  Bacchelli, and Premkumar Devanbu. 2016.
\newblock On the" naturalness" of buggy code.
\newblock In \emph{2016 IEEE/ACM 38th International Conference on Software
  Engineering (ICSE)}, pages 428--439. IEEE.

\bibitem[{Richter and Wehrheim(2022)}]{richter2022tssb}
Cedric Richter and Heike Wehrheim. 2022.
\newblock Tssb-3m: Mining single statement bugs at massive scale.
\newblock \emph{arXiv preprint arXiv:2201.12046}.

\bibitem[{Sennrich et~al.(2016)Sennrich, Haddow, and
  Birch}]{DBLP:conf/acl/SennrichHB16a}
Rico Sennrich, Barry Haddow, and Alexandra Birch. 2016.
\newblock Neural machine translation of rare words with subword units.
\newblock In \emph{{ACL} {(1)}}. The Association for Computer Linguistics.

\bibitem[{Spadini et~al.(2018)Spadini, Aniche, and Bacchelli}]{Spadini2018}
Davide Spadini, Maur\'{i}cio Aniche, and Alberto Bacchelli. 2018.
\newblock \href {https://doi.org/10.1145/3236024.3264598} {{PyDriller: Python
  framework for mining software repositories}}.
\newblock In \emph{Proceedings of the 2018 26th ACM Joint Meeting on European
  Software Engineering Conference and Symposium on the Foundations of Software
  Engineering - ESEC/FSE 2018}, pages 908--911, New York, New York, USA. ACM
  Press.

\bibitem[{Tufano et~al.(2018)Tufano, Watson, Bavota, Di~Penta, White, and
  Poshyvanyk}]{tufano2018empirical}
Michele Tufano, Cody Watson, Gabriele Bavota, Massimiliano Di~Penta, Martin
  White, and Denys Poshyvanyk. 2018.
\newblock An empirical investigation into learning bug-fixing patches in the
  wild via neural machine translation.
\newblock In \emph{Proceedings of the 33rd ACM/IEEE International Conference on
  Automated Software Engineering}, pages 832--837.

\bibitem[{Vasic et~al.(2019)Vasic, Kanade, Maniatis, Bieber, and
  Singh}]{vasic2019neural}
Marko Vasic, Aditya Kanade, Petros Maniatis, David Bieber, and Rishabh Singh.
  2019.
\newblock Neural program repair by jointly learning to localize and repair.
\newblock \emph{arXiv preprint arXiv:1904.01720}.

\bibitem[{Vaswani et~al.(2017)Vaswani, Shazeer, Parmar, Uszkoreit, Jones,
  Gomez, Kaiser, and Polosukhin}]{vaswani2017attention}
Ashish Vaswani, Noam Shazeer, Niki Parmar, Jakob Uszkoreit, Llion Jones,
  Aidan~N Gomez, {\L}ukasz Kaiser, and Illia Polosukhin. 2017.
\newblock Attention is all you need.
\newblock \emph{Advances in neural information processing systems}, 30.

\bibitem[{Vinyals et~al.(2015)Vinyals, Fortunato, and
  Jaitly}]{vinyals2015pointer}
Oriol Vinyals, Meire Fortunato, and Navdeep Jaitly. 2015.
\newblock Pointer networks.
\newblock \emph{Advances in neural information processing systems}, 28.

\bibitem[{Wang et~al.(2021)Wang, Wang, Joty, and Hoi}]{wang2021codet5}
Yue Wang, Weishi Wang, Shafiq~R. Joty, and Steven C.~H. Hoi. 2021.
\newblock Codet5: Identifier-aware unified pre-trained encoder-decoder models
  for code understanding and generation.
\newblock pages 8696--8708.

\bibitem[{Xie et~al.(2013)Xie, Chen, Kuo, and Xu}]{xie2013theoretical}
Xiaoyuan Xie, Tsong~Yueh Chen, Fei-Ching Kuo, and Baowen Xu. 2013.
\newblock A theoretical analysis of the risk evaluation formulas for
  spectrum-based fault localization.
\newblock \emph{ACM Transactions on Software Engineering and Methodology
  (TOSEM)}, 22(4):1--40.

\bibitem[{Yang et~al.(2016)Yang, Yang, Dyer, He, Smola, and
  Hovy}]{yang2016hierarchical}
Zichao Yang, Diyi Yang, Chris Dyer, Xiaodong He, Alex Smola, and Eduard Hovy.
  2016.
\newblock Hierarchical attention networks for document classification.
\newblock In \emph{Proceedings of the 2016 conference of the North American
  chapter of the association for computational linguistics: human language
  technologies}, pages 1480--1489.

\bibitem[{Yasunaga and Liang(2020)}]{yasunaga2020graph}
Michihiro Yasunaga and Percy Liang. 2020.
\newblock Graph-based, self-supervised program repair from diagnostic feedback.
\newblock In \emph{International Conference on Machine Learning}, pages
  10799--10808. PMLR.

\bibitem[{Yasunaga and Liang(2021)}]{yasunaga2021break}
Michihiro Yasunaga and Percy Liang. 2021.
\newblock Break-it-fix-it: Unsupervised learning for program repair.
\newblock In \emph{International Conference on Machine Learning}, pages
  11941--11952. PMLR.

\bibitem[{Zhang et~al.(2018)Zhang, Zhang, Harman, Hao, Jia, and
  Zhang}]{zhang2018predictive}
Jie Zhang, Lingming Zhang, Mark Harman, Dan Hao, Yue Jia, and Lu~Zhang. 2018.
\newblock Predictive mutation testing.
\newblock \emph{IEEE Transactions on Software Engineering}, 45(9):898--918.

\bibitem[{Zhu et~al.(2021)Zhu, Sun, Xiao, Zhang, Yuan, Xiong, and
  Zhang}]{zhu2021syntax}
Qihao Zhu, Zeyu Sun, Yuan-an Xiao, Wenjie Zhang, Kang Yuan, Yingfei Xiong, and
  Lu~Zhang. 2021.
\newblock A syntax-guided edit decoder for neural program repair.
\newblock In \emph{Proceedings of the 29th ACM Joint Meeting on European
  Software Engineering Conference and Symposium on the Foundations of Software
  Engineering}, pages 341--353.

\bibitem[{Zou et~al.(2019)Zou, Liang, Xiong, Ernst, and
  Zhang}]{zou2019empirical}
Daming Zou, Jingjing Liang, Yingfei Xiong, Michael~D Ernst, and Lu~Zhang. 2019.
\newblock An empirical study of fault localization families and their
  combinations.
\newblock \emph{IEEE Transactions on Software Engineering}, 47(2):332--347.

\end{thebibliography}

\end{document}